\documentclass{article}
\usepackage[dvips]{graphicx}
\usepackage{epsfig}
\usepackage{amssymb}
\usepackage{multicol}
\begin{document}
\begin{center}
{\LARGE \bf Flows in complex biochemical networks: Role of low degree nodes}
\end{center}
\begin{center}
\noindent \renewcommand{\thefootnote}{\fnsymbol{footnote}} {\bf
Areejit Samal$^{1,2}$ and Sanjay Jain$^{1,3,4}$ \footnote[1]{Email:
jain@physics.du.ac.in}}
\end{center}
\begin{center}
{\small \it
$^1$Department of Physics and Astrophysics, University of Delhi, Delhi 110007, India\\
$^2$Max Planck Institute for Mathematics in the Sciences, Inselstr. 22, 04103 Leipzig, Germany\\
$^3$Jawaharlal Nehru Centre for Advanced Scientific Research, Bangalore 560064, India\\
$^4$Santa Fe Institute, 1399 Hyde Park Road, Santa Fe, NM 87501, USA\\
}
\end{center}
\subsection*{Abstract}
Metabolic networks have two properties that are generally regarded
as unrelated: One, they have metabolic reactions whose single
knockout is lethal for the organism, and two, they have correlated
sets of reactions forming functional modules. In this review we
argue that both essentiality and modularity seem to arise as a
consequence of the same structural property: the existence of low
degree metabolites. This observation allows a prediction of (a)
essential metabolic reactions which are potential drug targets in
pathogenic microorganisms and (b) regulatory modules within
biological networks, from purely structural information about the
metabolic network.


\subsection*{Introduction}

In the course of evolution organisms have developed redundancies in
their intracellular networks so as to tolerate random failures such
as the knockout of a single gene or reaction. Nevertheless certain
genes or reactions are `essential' for growth and their single
knockout can render the cell unviable. A question arises: What makes
a gene or reaction essential? Is there some special structural
property characterizing essential reactions in metabolic networks?
Here we review some of our work which identifies such a property and
describes it in network terms. This work shows that most essential
metabolic reactions in {\it E. coli}, {\it S. cerevisiae} and {\it
S. aureus} can be explained by the fact that they are associated
with a low degree metabolite \cite{SSGKRJ2006}.

A second property of biological networks is their modularity
\cite{HHLM1999}, which is important, among other things, for their
robustness and evolvability. Modularity appears in several
biochemical networks, including metabolic networks and genetic
regulatory networks. One can ask: Are there structures at a lower
level of biological organization such as metabolism, perhaps
reflecting basic chemical constraints, that cause, or at least
encourage, modularity to emerge at a higher and evolutionarily more
flexible level of organization such as the genetic regulatory
network? It seems that metabolites with a low degree of connectivity
could provide one such structure. Low degree metabolites participate
in very few reactions in the network, which may be due to some
feature of their chemical structure that prohibits ready association
with other molecules. By virtue of their low degree these
metabolites also contribute to a rigidity or coherence of reaction
fluxes in the network resulting in clusters of highly correlated
reactions. It turns out that these reaction clusters correspond to
genetic regulatory modules with high probability, as captured in the
structure of operons in {\it E. coli} \cite{SSGKRJ2006}.

Thus we argue that low degree metabolites are implicated in two
distinct properties of biochemical networks, one, the essentiality
of certain metabolic reactions, and two, the modularity of genome
organization.


\subsection*{Relationship between essential reactions and low degree metabolites}

\subsubsection*{Essential metabolic reactions: Definition and in-silico identification}

A metabolic reaction is designated as `essential' for an organism in
a given environmental condition if its single knockout from the
network renders the organism unviable in that condition. Essential
reactions are therefore critical for the survival of the organism;
as such their very existence represents fragilities associated with
metabolism. The experimental determination of all essential genes
and associated reactions in an organism under different
environmental conditions is extremely tedious and time consuming.
Alternatively, in-silico methods such as flux balance analysis (FBA)
can be used for fast prediction of all essential metabolic reactions
in an organism under different environmental conditions.

FBA is a computational modelling technique which can be used to
obtain the optimal growth rate of the organism and the steady state
fluxes of all reactions in the metabolic network for any given
growth medium (for a review see ref. \cite{PRP2004}). Using FBA, a
reaction is said to be `essential' for a particular medium, if
switching the reaction off results in a zero optimal growth rate for
that medium. In \cite{SSGKRJ2006}, FBA was used to compute essential
reactions in the {\it E. coli} metabolic network (version iJR904
\cite{RVSP2003}) under 89 different aerobic minimal media. These 89
media correspond to all possible distinct single organic sources of
carbon under which {\it E. coli} can grow in aerobic conditions, as
determined using FBA. It was found that the number of essential
reactions varied between 200 and 240 across the 89 minimal media. A
reaction was designated as `globally essential' for {\it E. coli},
if it was essential for all the 89 media. The number of globally
essential reactions was obtained to be 164 for the {\it E. coli}
metabolic network. We now turn to an explanation, in terms of the
underlying structure of the {\it E. coli} metabolic network, as to
why these reactions are essential.

\subsubsection*{Uniquely produced and uniquely consumed metabolites}

The metabolic network may be represented as a bipartite graph
consisting of two types of nodes: metabolites and reactions. In a
directed bipartite graph, there are two types of links: (a) from
metabolite nodes to reaction nodes and (b) from reaction nodes to
metabolite nodes. A metabolite is designated as `uniquely produced'
or `UP' (`uniquely consumed' or `UC'), if there is only a single
reaction in the metabolic network that produces (consumes) the
metabolite \cite{SSGKRJ2006}. A UP(UC) metabolite has in-degree
(out-degree) equal to unity in the bipartite graph. A metabolite
that is both UP and UC is designated as a `UP-UC metabolite'. In
general, a metabolite that is either UP or UC or both has a low
degree in the metabolic network as it participates in very few
reactions. Similarly, a reaction is designated as `uniquely
producing' or `UP' (`uniquely consuming' or `UC'), if it produces
(consumes) a UP(UC) metabolite in the bipartite metabolic network.
Reactions in the metabolic network that are either UP or UC or both
are designated as `UP/UC reactions'. Examples of UP(UC) metabolites
and reactions in the {\it E. coli} metabolic network are shown in
Fig. \ref{cluster}.

\subsubsection*{Almost all essential reactions are UP/UC}

The {\it E. coli} metabolic network database iJR904 which contains
761 metabolites participating in 931 reactions was studied. Starting
from the list of 931 reactions in the {\it E. coli} network, a
transformed network containing 1176 reactions was constructed by
replacing each reversible reaction with two irreversible reactions
(one each for the forward and backward direction). In the
reconstructed metabolic network databases discussed here, there are
certain reactions that can only have a zero flux value under any
steady state. Such reactions have been referred to as `blocked'
reactions' \cite{SS1991,BNSM2004}, and their occurrence points
towards the incompleteness of metabolic network databases. 290
blocked reactions were found in the {\it E. coli} metabolic network;
these were removed from the transformed network of 1176 reactions to
obtain a reduced network of 886 reactions representing {\it E. coli}
metabolism. The removal of blocked reactions from the original
network does not affect the allowed steady state fluxes through the
rest of the network, and in particular does not alter the set of
essential reactions obtained using FBA. The number of UP(UC)
reactions in the reduced metabolic network of {\it E. coli} was
found to be 245 (218). The number of reactions that were either UP
or UC or both (the `UP/UC reactions') in the reduced metabolic
network of {\it E. coli} was 352.

\begin{figure}
\centering
\includegraphics[height=10cm]{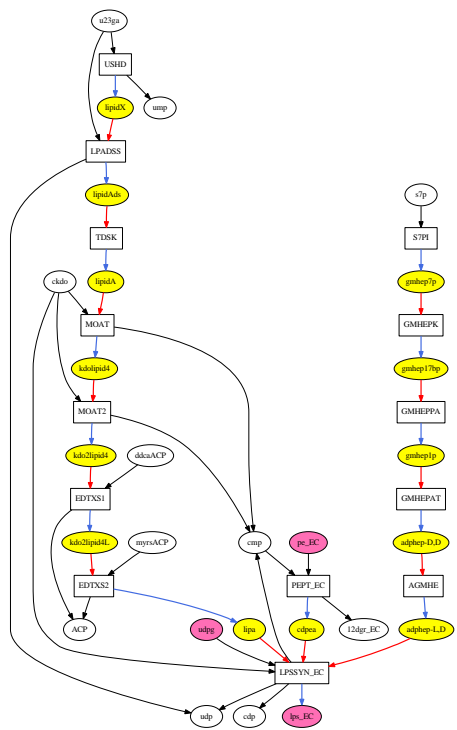}
\caption{UP-UC metabolites in the {\it E. coli} metabolic network
forming a UP-UC cluster. The 14 reactions that belong to the UP-UC
cluster are `globally essential'. Rectangles represent reactions and
ovals metabolites. Yellow ovals represent UP-UC metabolites and pink
ovals represent biomass metabolites. A blue (red) link represents
the production (consumption) of a UP (UC) metabolite; otherwise the
link is shown in black. To reduce clutter, nodes corresponding to
$adp$, $atp$, $h$, $h2o$, $pi$ and $ppi$ have been omitted.
Abbreviation of metabolite and reaction names are as in
\cite{RVSP2003}.} \label{cluster}
\end{figure}

It is evident that if a UP (or UC) metabolite is an essential
intermediate in the production of a metabolite that is part of the
biomass reaction contributing to the growth of the organism then the
reaction responsible for the production (or consumption) of that UP
(or UC) metabolite becomes essential for the growth of the organism.
However, the converse is not obvious, i.e., there is no reason why
all or most essential reactions should be UP or UC; there could, in
principle, be other reasons why a reaction is essential. Of the 164
globally essential reactions in the {\it E. coli} metabolic network,
156 were either UP or UC in the reduced network. This explains why
the subset of 156 reactions are globally essential in {\it E. coli}.
By definition, there exists some metabolite (that is presumably
required for generating biomass) whose production or consumption can
only be performed by these reactions, thereby making them essential.
Further, we found the probability of such a high overlap between the
set of globally essential reactions and set of UP/UC reactions
occurring by pure chance is very small (see the $p$-value in Table
\ref{essential}). In Fig. \ref{cluster}, we have shown 14 UP/UC
reactions that are globally essential reactions in the {\it E. coli}
metabolic network; they are required for the production of the
biomass metabolite lipopolysaccharide (abbreviated as lps\_EC in
Fig. \ref{cluster}). We also determined globally essential reactions
in the metabolic networks of two other organisms, {\it S.
cerevisiae} (version iND750 \cite{DHP2004}) and {\it S. aureus}
(version iSB619 \cite{BP2005}), and again found almost all globally
essential reactions in the two organisms to be either UP or UC in
their respective reduced networks (see Table \ref{essential}).

Mahadevan and Palsson \cite{MP2005}\ had earlier observed that the
lethality fraction (the fraction of reactions a metabolite is
involved in that are essential) of the low degree metabolites is on
average comparable to high degree metabolites. However, they did not
realize that the essential reactions are explained by their
association with the low degree metabolites rather than the high
degree ones. Our work has led to the understanding that the
essential reactions may involve other metabolites of higher degree,
but their essentiality is due to their uniqueness in producing or
consuming a UP or UC metabolite \cite{SSGKRJ2006}.

It was also found that of the 352 UP/UC reactions in the {\it E.
coli} reduced metabolic network 288 were essential in at least one
of the 89 aerobic minimal media. The intersection of the so-called
high-flux backbone reactions \cite{AKVOB2004} across these media
also consists primarily of UP/UC reactions (116 out of 124). There
exist multiple flux vectors (or multiple solutions of FBA) with the
same optimal growth rate in any given medium \cite{RP2004}. This
multiplicity represents the redundancy in metabolic pathways. The
high-flux backbone changes across the multiple solutions. The part
of the high-flux backbone that is conserved across the multiple
solutions also consists primarily of the UP/UC reactions (160 out of
197 in glucose minimal medium) \cite{S2009}. These facts underline
the importance of UP/UC reactions in maintaining flows across
metabolic networks.

\begin{table}
\centering
\begin{tabular}{|l|l|l|l|}
\hline
{\bf \small Organism} & {\bf \small \it E. coli}&{\bf \small \it S. cerevisiae} & {\bf \small \it S. aureus}\\
\hline
{\small Number of reactions in the transformed network} & {\small 1176}&{\small 1579} & {\small 865}\\
\hline
{\small Number of reactions in the reduced network} & {\small 886} & {\small 779} & {\small 571}\\
\hline
{\small Number of globally essential reactions} & {\small 164} & {\small 127} & {\small 196}\\
\hline
{\small Number of globally essential reactions} & {\small 156} & {\small 117} & {\small 182 }\\
{\small that are UP or UC} & {\tiny {($p<10^{-62}$)}} & {\tiny {($p<10^{-41}$)}} & {\tiny {($p<10^{-58}$)}}\\
\hline
\end{tabular}
\caption{Comparison of globally essential reactions and UP/UC
reactions in the metabolic networks of {\it E. coli}, {\it S.
cerevisiae} and {\it S. aureus}. This table is adapted from Ref.
\cite{SSGKRJ2006}} \label{essential}
\end{table}

\subsection*{UP-UC clusters predict regulatory modules and are network motifs}

A UP-UC metabolite has one reaction that produces it and one
reaction that consumes it in the metabolic network. A steady state
is defined as one where all metabolite concentrations and reaction
velocities are constant. In any steady state, the flux of the
reaction producing a UP-UC metabolite is always proportional to the
flux of the reaction consuming the metabolite. A `UP-UC cluster' of
reactions was defined as a set of reactions connected by UP-UC
metabolites \cite{SSGKRJ2006}. An example of a UP-UC cluster of
reactions in the reduced  metabolic network of {\it E. coli} is
shown in Fig. \ref{cluster}. In steady state, fluxes of all
reactions that are part of a single UP-UC cluster are proportional
to each other and fixing the flux of any reaction in a UP-UC cluster
fixes the fluxes of all other reactions in the cluster under steady
state. UP-UC clusters of reactions are special cases of
reaction/enzyme subsets \cite{PSNMS1999}, co-sets \cite{RP2004} and
fully coupled reactions \cite{BNSM2004} that have been discussed
earlier in the literature. Further, for any steady state analysis,
each UP-UC cluster can be replaced by a single effective reaction,
and this can be used to coarse-grain metabolic networks
\cite{PSNMS1999}.

Since the fluxes of reactions forming a UP-UC cluster have fixed
ratios with respect to each other for all steady states, the set of
genes that code for enzymes catalyzing various reactions of the
cluster may be expected to be coregulated forming a transcriptional
module. In prokaryotes genes are grouped into transcriptional
modules called operons. The set of genes that form a single operon
are always transcribed together. We have investigated whether the
genes coding for enzymes catalyzing reactions of a UP-UC cluster are
part of the same operon in {\it E. coli}. It was found
\cite{SSGKRJ2006} that from the set of 251 genes corresponding to
reactions of all the UP-UC clusters, two genes belonging to the same
UP-UC cluster have a much greater probability (more than 50 times)
of lying on the same operon than a randomly chosen pair (the
probabilities are 0.29 and 0.0057, respectively). Thus, the set of
genes that correspond to a UP-UC cluster in the {\it E. coli}
metabolic network are strongly correlated with regulatory modules at
the genetic level.

The bunching up of UP-UC metabolites next to each other in the
metabolic network results in the formation of UP-UC clusters with
more than two reactions. A natural question arises: Is it expected
for a network like the {\it E. coli} metabolic network to have many
large UP-UC clusters? This question may be answered by comparing
with a null model. The distribution of UP-UC clusters in the real
{\it E. coli} metabolic network was compared with a suitably
randomized version of the original network. It was found that the
real metabolic network of {\it E. coli} has its UP-UC metabolites
bunched up next to each other, forming larger clusters than may be
expected in random networks with the same local connectivity
properties as the original network \cite{SSGKRJ2006}. `Network
motifs' have been defined as patterns of interconnections that occur
in different parts of a network at frequencies much higher than
those found in randomized networks \cite{MSIKCA2002}. Thus, larger
size UP-UC clusters may be collectively considered as analogous to a
network motif, since they are over-represented in the real {\it E.
coli} metabolic network. The larger UP-UC clusters in the real
network may facilitate the regulation of certain metabolic pathways
inside the organism.

\subsection*{Discussion}
Metabolic networks inside different organisms have been shown to
follow a power law degree distribution  \cite{JTAOB2000,WF2001}
which is characterized by the presence of high degree metabolites
referred to as `hubs'. It has been suggested that one of the
important consequences of a power law degree distribution is the
vulnerability of the network to selective attack on hubs while being
robust to random deletion of nodes from the network \cite{AJB2000}.
For the protein-protein interaction network of {\it S. cerevisiae},
it was shown that the essentiality of a protein is correlated with
its degree in the network \cite{JMBO2001}. This observation has been
suggested as evidence for the importance of the hubs in maintaining
the overall structure and function of cellular networks. Although
the role of high degree metabolites or hubs in maintaining the
overall structure of the metabolic networks has been well emphasized
in the literature, the role of low degree metabolites has attracted
little or no attention. In our study, we find that certain low
degree metabolites explain globally essential reactions and
introduce fragility for flows in metabolic network. We have further
shown that the low degree metabolites as opposed to high degree
metabolites explain essential reactions in metabolic networks
\cite{SSGKRJ2006}. Thus, it is the low degree metabolites that are
critical from the point of view of functional robustness of the
metabolic system.

The fragility caused by low degree metabolites in metabolic networks
can have potential applications in medicine. In our work, we have
observed that essential reactions are explained by UP/UC structure in
three organisms. Thus, it is likely that UP/UC structure explains
essential reactions in other organisms that are pathogens for humans.
This generates candidate targets for therapeutic intervention. It is
conceivable that drugs could be found that incapacitate the enzymes
catalyzing essential reactions in pathogens.

The above work highlights the essentiality of reactions associated
with low degree metabolites. It also shows that these metabolites
lead to correlated clusters of reactions in the metabolic network
that preferentially correspond to genetic regulatory modules in {\it
E. coli}. Thus the same structural property the existence of low
degree metabolites seems to be associated with two seemingly
unrelated properties of the system: essentiality and modularity.
While the essentiality of a reaction is obviously connected with a
fragility of the organism (knockout of the reaction is lethal),
modularity is generally believed to contribute to the organism's
robustness and evolvability. This might be an example of a rather
general property of complex systems, that their robustness in some
dimension often goes hand in hand with fragility in another
dimension \cite{CD2002}.


\subsubsection*{Acknowledgement}

{\small We would like to thank Shalini Singh, Varun Giri, Sandeep
Krishna and N. Raghuram for collaboration. SJ acknowledges financial
support for his research program through grants from the Department
of Biotechnology, Government of India.}

\small{

}

\end{document}